\documentclass{article} 
\usepackage{iclr2026_conference,times}

\usepackage{amsmath,amsfonts,bm}

\def\eqref#1{equation~\ref{#1}}

\def\1{\bm{1}}

\DeclareMathAlphabet{\mathsfit}{\encodingdefault}{\sfdefault}{m}{sl}
\SetMathAlphabet{\mathsfit}{bold}{\encodingdefault}{\sfdefault}{bx}{n}

\usepackage{hyperref}
\usepackage{makecell} 
\usepackage{url}
\usepackage{booktabs}
\usepackage{multirow}
\usepackage{xcolor}
\usepackage{graphicx}
\usepackage{xcolor}
\usepackage{colortbl}
\usepackage{algorithm}
\usepackage{algorithmic}
\usepackage{amsmath}
\usepackage{amssymb}
\usepackage{amsfonts}
\usepackage{float}
\usepackage{enumitem}
\usepackage{wrapfig}
\usepackage{array}
\usepackage{longtable}
\usepackage[table]{xcolor} 
\usepackage{makecell}

\usepackage{subcaption}
\usepackage{bbm}

\title{RiboPO: Preference Optimization for \\ Structure- and Stability-Aware RNA Design}

\iclrfinalcopy

\author{%
\begin{tabular}{l} 
Minghao Sun\textsuperscript{1}\thanks{Equal contribution.} \quad
Hanqun Cao\textsuperscript{2}\footnotemark[1] \quad
Zhou Zhang\textsuperscript{4}\footnotemark[1] \quad
Chen Wei\textsuperscript{1} \quad 
Liang Wang\textsuperscript{1,5} \\
Tianrui Jia\textsuperscript{1} \quad
Zhiyuan Liu\textsuperscript{1} \quad
Tianfan Fu\textsuperscript{4} \quad
Xiangru Tang\textsuperscript{6} \quad
Yejin Choi\textsuperscript{3} \\
Pheng\mbox{-}Ann Heng\textsuperscript{2} \quad
Fang Wu\textsuperscript{3}\thanks{Corresponding author.} \quad
Yang Zhang\textsuperscript{1}\footnotemark[2]
\end{tabular}
\\[6pt] 
\textsuperscript{1}National University of Singapore \quad
\textsuperscript{2}The Chinese University of Hong Kong \\
\textsuperscript{3}Stanford University \quad
\textsuperscript{4}Nanjing University \quad
\textsuperscript{5}Chinese Academy of Sciences \\
\textsuperscript{6}Yale University 
}

\begin{document}

\maketitle

\begin{abstract}
Designing RNA sequences that reliably adopt specified three-dimensional structures while maintaining thermodynamic stability remains challenging for synthetic biology and therapeutics. Current inverse folding approaches optimize for sequence recovery or single structural metrics, failing to simultaneously ensure global geometry, local accuracy, and ensemble stability—three interdependent requirements for functional RNA design. This gap becomes critical when designed sequences encounter dynamic biological environments.
We introduce \textbf{RiboPO}, a \textbf{Ribo}nucleic acid \textbf{P}reference \textbf{O}ptimization framework that addresses this multi-objective challenge through \emph{reinforcement learning from physical feedback} (RLPF). RiboPO fine-tunes gRNAde by \emph{constructing preference pairs from composite physical criteria} that couple global 3D fidelity and thermodynamic stability. Preferences are formed using structural gates, \emph{pLDDT} geometry assessments, and thermostability proxies with variability-aware margins, and the policy is updated with Direct Preference Optimization (DPO). On RNA inverse folding benchmarks, RiboPO demonstrates a superior balance of structural accuracy and stability. Compared to the best non-overlap baselines, our multi-round model improves Minimum Free Energy (MFE) by \textbf{12.3\%} and increases secondary-structure self-consistency (EternaFold scMCC) by \textbf{20\%}, while maintaining competitive 3D quality and high sequence diversity. In sampling efficiency, RiboPO achieves \textbf{11\% higher pass@64} than the gRNAde base under the conjunction of multiple requirements. A multi-round variant with preference-pair reconstruction delivers additional gains on unseen RNA structures. These results establish RLPF as an effective paradigm for structure-accurate and ensemble-robust RNA design, providing a foundation for extending to complex biological objectives.
\end{abstract}

\section{Introduction}

The inverse folding problem in RNA design addresses a fundamental computational challenge: determining which nucleotide sequence will reliably fold into a specified three-dimensional backbone structure \citep{tang2024survey,tan2025r3design}. This problem represents the cornerstone of rational RNA engineering \citep{wong2024deep}, enabling systematic design of functional molecules for gene regulation \citep{green2014toehold}, therapeutics \citep{yip2024therapeutic}, and synthetic biology applications \citep{pfeifer2023harnessing}.

\emph{Designability}—the number/density of sequences that realize a target fold—matters because highly designable structures are easier to discover, more mutation-robust \citep{england2003naturalselectdesignablefold}, and tend to inhabit thermodynamically favorable basins \citep{designability_li_1996}. Optimizing geometry alone may reach thin, fragile ridges in sequence space; coupling geometry with ensemble stability biases search toward more designable regions \citep{schuster1994rna2dshape}.

Contemporary geometric neural networks have achieved significant progress on this task \citep{wu2023integration,wu2023hierarchical,liu2025sentences,patil2024towards}. Models such as gRNAde \citep{joshi2025grnade} utilize graph neural networks with geometric vector perceptrons to capture spatial relationships within RNA tertiary structures, delivering impressive sequence recovery performance across benchmark datasets \citep{ganser2019roles}.
However, RNA molecules exhibit inherent conformational flexibility that causes them to sample multiple competing structures in solution, unlike proteins that maintain relatively stable native folds \citep{zadeh2011nucleic}.

Current inverse folding methods \citep{Hou2025ridiffusion, wong2024deep}, when optimized for sequence similarity or a single structural proxy, often neglect \emph{designability} and thermodynamic stability, and therefore \emph{do not} consistently adopt the intended conformation under physiological conditions.

Prevailing evaluations emphasize geometric correspondence (e.g., TMscore \citep{zhang2004tmscore} and RMSD for global alignment, pLDDT \citep{jumper2021alphafold2} for local fidelity), but these alone provide limited insight into \emph{ensemble viability} and \emph{designability}. As a result, sequences can score well structurally yet occupy narrow, unstable basins that admit competing conformations in solution \citep{bernard2024rnadvisor}. This evaluation-centric limitation produces sequences that achieve excellent structural scores while exhibiting poor ensemble behavior, sampling multiple competing conformations, and failing to demonstrate practical utility in biological applications \citep{zhou2023rna}. From a design perspective, single-objective optimization cannot fully address the multifaceted challenges inherent to RNA structural stability, where geometric fidelity must be balanced against thermodynamic preferences and conformational robustness.
\begin{figure}[!t] 
    \centering 
    \includegraphics[width=0.9\textwidth]{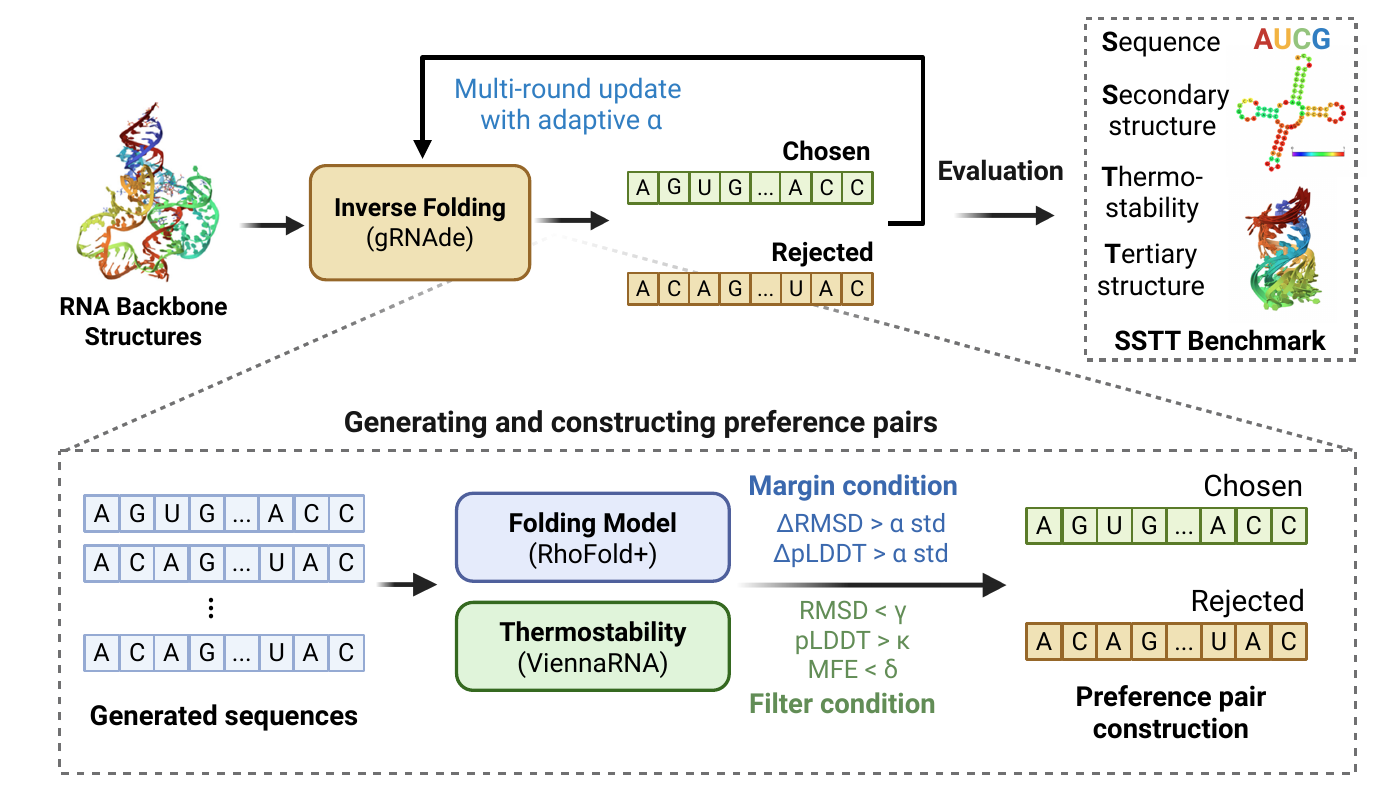} 
    \vspace{-1em}
    \caption{Overview of RiboPO framework.} 
    \label{fig:main} 
\end{figure}

To address these limitations, we present \textbf{RiboPO}, a reinforcement-learning–from–physical-feedback framework for RNA tertiary inverse folding that \emph{explicitly} incorporates thermodynamic stability alongside geometry. Our approach reframes RNA inverse folding as a multi-objective preference optimization problem, representing a paradigm shift from purely geometric to biophysically-informed design. Unlike previous RL applications limited to RNA secondary structure prediction, RiboPO operates on full tertiary structures while balancing structural accuracy with thermodynamic stability.
Our contributions are threefold:
\begin{itemize}[leftmargin=*]
\item  We incorporate thermodynamic stability measures into the reward for RNA inverse folding, representing the first systematic approach to include ensemble stability properties that existing RNA design methods have consistently overlooked.
Our framework employs offline RL with precomputed training pairs, enabling efficient use of costly biophysical calculations during training.
\item Second, our multi-objective preference optimization framework enables iterative refinement across competing design objectives through successive training rounds. This approach allows dynamic balancing between structural fidelity, local interaction accuracy, and thermodynamic stability without requiring online structure prediction during training.
\item Third, we establish the \textbf{SSTT Benchmark} (\textbf{S}equence, \textbf{S}econdary, \textbf{T}ertiary, \textbf{T}hermostability) for comprehensive evaluation of RNA sequence generation quality. Our analysis reveals fundamental limitations in current geometric-only approaches and demonstrates that preference-based optimization significantly improves both designability and thermodynamic stability while maintaining competitive sequence recovery performance.
\end{itemize}

On the DAS split \citep{das2010farfar, joshi2025grnade} with our comprehensive SSTT suite, RiboPO achieves \emph{general} improvements across dimensions: secondary-structure self-consistency (EternaFold scMCC: $+0.12$, \textbf{+20\%}; $0.72$ vs.\ $0.60$) and thermostability (MFE: \textbf{$-4.03$}~kcal/mol, \textbf{12.3\%}; $-36.86$ vs.\ $-32.83$), while maintaining competitive recovery and strong tertiary fidelity (RMSD \textbf{10.23}\AA\ vs.\ 10.66\AA; \%RMSD$\le 8$\AA: \textbf{0.51} vs.\ 0.46). In \S\ref{sec:experiments}.\ref{subsec:benchmark_analysis} we show that these shifts correspond to movement into more \emph{designable} basins. In \S\ref{sec:experiments}.\ref{subsec:ablation_studies}, ablations validate the necessity of (i) variability-aware pair construction, (ii) DPO+$\lambda_{\text{SFT}}$ loss, and (iii) the multi-round schedule. Practically, sampling efficiency improves substantially: 

Overall, preference-based optimization with physical feedback yields sequences that are not only structurally accurate but also ensemble-robust—advancing RNA design toward practically viable, mutation-tolerant solutions.


\section{Related Work}
\label{sec:related}

\paragraph{RNA structure prediction and inverse folding.}
Recent models have advanced RNA structure prediction across tertiary and secondary levels. For tertiary structure, traditional methods like DRFold \citep{li2023integrating}  and DeepFoldRNA \citep{pearce2022novo} combine primary/secondary structure information with energy functions for prediction and optimization. AlphaFold2-inspired approaches including NuFold \citep{kagaya2025nufold}, trRoseTTARNA \citep{wang2023trrosettarna} and RhoFold+ \citep{shen2024accurate} employ EvoFormer architectures to predict structures from sequence and secondary structure data \citep{jumper2021alphafold2}. RoseTTAFoldNA \citep{baek2024accurate} and AlphaFold 3 \citep{abramson2024accurate} extend prediction capabilities to protein-nucleic acid complexes and multi-component assemblies. For secondary structure, traditional methods RNAfold/ViennaRNA \citep{lorenz2011viennarna} calculate partition functions and base-pairing probabilities through minimum free energy optimization. SPOT-RNA \citep{singh2019rna} uses 2D CNNs with transfer learning for non-canonical base pairs, E2EFold \citep{chen2020rna} predicts contact matrices, and RFold \citep{tan2024deciphering} employs bidirectional decomposition to constrain cross-pairing interactions. Current inverse folding methods predominantly use GNN-based structural encoders, including RNAinformer \citep{patil2024towards}, R3Design \citep{tan2025r3design}, RDesign \citep{tanrdesign}, RhoDesign \citep{wong2024deep}, RiFold \citep{liu2025sentences}, AlignIF \citep{wang2025alignment}, and gRNAde \cite{joshi2025grnade}. RiboDiffusion \citep{huang2024ribodiffusion} alternatively employs structure-conditioned discrete diffusion models. Notable advances include RhoDesign's \cite{wong2024deep} experimental validation, RiFold's language modeling framework \citep{liu2025sentences}, and gRNAde's \citep{joshi2025grnade} extension from single-state to multi-state design.

\paragraph{Reinforcement Learning and Preference Optimization for Inverse Folding.}
RL and preference optimization are increasingly used to steer inverse folding toward structural fidelity and stability. ProteinZero explores online RL for self-improvement with proxy rewards \citep{wang2025proteinzero}, while ResiDPO introduces residue-level Direct Preference Optimization (DPO) using structure-based rewards to improve designability \citep{xue2025improving}. Diversity-regularized DPO has been applied to peptide inverse folding \citep{park2024improving}. PLM-RL combined RL with the Protein Language Models \citep{cao2025supervisionexplorationdoesprotein}. RL has also been combined with diffusion models: RL-DIF optimizes structure-conditioned categorical diffusion \citep{ektefaie2024reinforcement}, and DRAKES backpropagates rewards through discrete diffusion trajectories for DNA/protein design \citep{wang2024fine}. Search-based strategies like ProtInvTree use reward-guided tree search \citep{liu2025protinvtree}, and lightweight, gradient-free finetuning for discrete sequences is explored in GLID$^2$E \citep{cao2025glid}. Finally, EnerBridge-DPO integrates energy-based preferences via Markov bridges with DPO to bias designs toward low-energy sequences \citep{rong2025enerbridge}. To the best of our knowledge, none of existing methods has demonstrated the effectiveness of RLPF on RNA.
\section{Methods}
\label{sec:method}

\subsection{Problem Formulation and Backbone-Conditioned Approach}
Ribonucleic acid (RNA) is a nucleic acid present in all living cells. An RNA molecule, often single-stranded, has a backbone made of alternating phosphate groups and sugar ribose. RNA has four nucleotides (also known as bases): adenine (A), guanine (G), cytosine (C), and uracil (U).

We formulate RNA inverse folding as a multi-objective optimization problem over the discrete RNA sequence space $\mathcal{S}=\{A,U,C,G\}^L$ ($L$ is the RNA sequence length), where we seek sequences that jointly maximize structural fidelity and thermodynamic stability:
\begin{equation}
s^*=\arg\max_{s\in\mathcal{S}}\left[f_{\text{struct}}(s,\mathcal{G})+f_{\text{thermo}}(s)\right],
\end{equation}
where $\mathcal{G}$ denotes the target backbone geometry, $f_{\text{struct}}$ captures structural quality metrics (e.g., pLDDT, RMSD), and $f_{\text{thermo}}$ quantifies thermodynamic favorability via free energy.

To solve this, we use a backbone-conditioned policy that generates sequences given $\mathcal{G}$. Specifically, our policy $\pi_\theta(s\mid\mathcal{G})$ follows the GNN-GVP architecture from gRNAde \citep{joshi2025grnade}, encoding the 3D backbone and producing sequences autoregressively:
\begin{equation}
\pi_\theta(s\mid\mathcal{G})=\prod_{i=1}^L \pi_\theta(s_i \mid s_{<i}, \mathcal{G}).
\end{equation}
We maintain a frozen reference policy $\pi_{\mathrm{ref}}$ initialized from the pre-trained gRNAde model to provide a stable baseline and regularize distributional drift.

Optimization proceeds via multi-round Direct Preference Optimization, which progressively shifts $\pi_\theta$ toward \emph{more optimized solutions}. After each round, the improved policy supplies higher-quality preference pairs for the next round, yielding an iterative curriculum that stabilizes training while aligning the generation distribution with the multi-objective targets above.

\subsection{Multi-Round Preference Construction}

\paragraph{Progressive Refinement Strategy.}
Our iterative approach leverages improved policies from previous rounds to generate candidate sequences in previously unexplored regions of sequence space \citep{belanger2019biological}. This addresses the exploration-exploitation trade-off inherent in RNA sequence optimization, where complex structure-function relationships create numerous local optima \citep{hofacker2010barmap,jimenez2013comprehensive}. Single-round optimization frequently converges to sequences with adequate structural similarity but suboptimal thermodynamic properties or limited designability under physiological conditions. By progressively refining discrimination criteria across rounds, our method identifies increasingly subtle quality differences that drive convergence toward globally optimal solutions, balancing structural accuracy with biophysical stability \citep{gonzalez2017preferential}. 
We operationalize this progressive refinement through round-wise tightening of preference criteria, as detailed below.

\paragraph{Preference Criteria.}
We establish high-quality preference pairs using systematic statistical thresholds tailored to the RNA inverse folding task. Given two sequences $s^w$ (winner/chosen) and $s^l$ (loser/rejected), a preference is recorded only when both filter and margin conditions are satisfied:

\begin{align}
\text{\textbf{Filter:}} \quad & \text{pLDDT}(g(s^w)) > 0.70 \land \text{RMSD}(\mathcal{G}, g(s^w)) < 8\,\text{\AA} \\
\text{\textbf{Margin:}} \quad & \underbrace{\left(\bigwedge_{m\in\{\text{pLDDT, RMSD}\}}|m(s^w)-m(s^l)|>\gamma_r\cdot\sigma_m\right)}_{\text{Structural Margin}} \land \underbrace{\left(\text{MFE}(s^w) < \text{MFE}(s^l)\right)}_{\text{Thermodynamic Preference}}
\end{align}

where $g$ denotes RNA folding approaches (RhoFold+ \citep{shen2024accurate} in this work). $\sigma_m$ is the standard deviation of metric $m(\cdot)$ estimated from the current candidate pool. These criteria exclude candidates unlikely to fold correctly, ensuring that preference labels are assigned within a biologically realistic design space and reducing downstream noise. The statistical margin $\gamma_r\sigma_m$ enforces the \emph{significance} of the structural preferences, effectively improving the signal-to-noise ratio of the implicit reward; normalization by $\sigma_m$ renders metrics on different scales commensurate and stabilizes the DPO likelihood-ratio updates.

Concretely, we implement this progressive refinement of structure and energy by decreasing $\gamma_r$ across iterations. Early rounds use looser margins to rapidly establish broad foldability; later rounds tighten the margins to capture subtle, near–native improvements (e.g., local geometry). This round-wise annealing functions as Curriculum Learning (CL): large early gaps yield high-confidence preferences that reduce label noise and reward sparsity, while smaller later gaps provide denser, finer-grained gradient signals. The schedule mirrors an exploration-to-exploitation transition and implicitly enforces a trust-region constraint relative to the reference policy, stabilizing updates of large autoregressive models.

\subsection{Reinforcement Learning from Physical Feedback via DPO}
\label{subsec:rlpf_dpo}
We construct preference pairs $(s^w, s^l)$ by sampling candidate sequences from the current policy $\pi_\theta(\,\cdot\,\mid\mathcal{G})$ for each target RNA backbone $\mathcal{G}$ and then applying the filter and margin criteria described above. This procedure ensures that $s^w$ and $s^l$ reflect biologically realistic, thermodynamically feasible regions of sequence space.

Our objective maximizes the preference likelihood ratio between $s^w$ and $s^l$ while anchoring against the reference policy:
\begin{equation}
\mathcal{L}_{\mathrm{DPO}}
= -\mathbb{E}\!\left[
\log \sigma \!\Bigl(
\beta \log \frac{\pi_\theta(s^w \mid \mathcal{G})}{\pi_\theta(s^l \mid \mathcal{G})}
- \beta \log \frac{\pi_{\mathrm{ref}}(s^w \mid \mathcal{G})}{\pi_{\mathrm{ref}}(s^l \mid \mathcal{G})}
\Bigr)
\right].
\end{equation}
where it directly shapes the policy toward generating sequences with better structural fidelity and thermodynamic stability. 
while the reference-policy term acts as a trust region to prevent destabilizing updates, preventing regressions toward suboptimal folds.

To stabilize training further, we add supervised fine-tuning (SFT) loss on the preferred sequences:
\begin{equation}
\mathcal{L}
= \mathcal{L}_{\mathrm{DPO}}
+ \lambda_{\mathrm{SFT}}\,
\mathbb{E}_{s^w}\bigl[-\log \pi_\theta(s^w \mid \mathcal{G})\bigr],
\end{equation}
where $\lambda_{\mathrm{SFT}}$ is a hyperparameter controlling the strength of this term.  
The SFT component acts as an explicit maximum-likelihood ``anchor" on sequences with higher designability and thermostability potential, reducing catastrophic drift and mode collapse often seen in RL-based fine-tuning.  


\subsection{SSTT Benchmark: Comprehensive Evaluation Framework}
\label{subsec:sstt}

Unlike existing RNA inverse folding benchmarks, which mainly evaluate sequence recovery and RMSD, we introduce the SSTT benchmark to assess performance across four complementary dimensions. This framework integrates thermodynamic stability and ensemble properties alongside traditional structural metrics, offering a multidimensional view of design quality. Detailed calculation procedures are provided in Appendix~\ref{subsec:sstt_detail}. {\textbf{S}equence Axis} captures how closely designed sequences resemble native ones and how broadly they explore sequence space. We assess recovery rate \citep{dauparas2022proteinmpnn} to measure native residue similarity, and 3-mer diversity \citep{shannon1948mathematical,Bokulich2024diversitykmer} to quantify sequence exploration capacity. {\textbf{S}econdary Structure Axis} evaluates whether designed sequences fold back into their intended secondary structures. We measure forward-folding self-consistency using Matthews Correlation Coefficient \citep{matthews1975comparison,mccmoreaccurate} on EternaFold \citep{waymentsteele2022eternafold} predictions, capturing base-pairing fidelity. {\textbf{T}ertiary Structure Axis} assesses three-dimensional fidelity at both global and local levels as well as steric and torsional realism. We evaluate TM-score \citep{zhang2004tmscore}, GDT, and RMSD for global similarity; pLDDT \citep{jumper2021alphafold2} for confidence; lDDT for local structure preservation; interaction network fidelity \citep{parisien2009newmetricsinf} for base-pairing accuracy; clash score \citep{word1999clashscore,davis2007molprobity} for geometric feasibility; and backbone torsion quality \citep{ramachandran1963stereochemistry} for stereochemical realism. {\textbf{T}hermostability Axis} quantifies the thermodynamic robustness of designed sequences under physiological conditions. We incorporate minimum free energy \citep{zuker1981optimal} for stability, ensemble defect \citep{zadeh2011nucleic} to measure target structure specificity, target structure probability \citep{McCaskill1990TheEquilibrium} to quantify conformational preference, Shannon entropy for ensemble sharpness, and melting temperature for thermal resilience. 
 
\section{Experiments \& Results} 
\label{sec:experiments}

\subsection{Experimental settings} 
\label{subsec:exp_settings}

\paragraph{Data Processing}
Our experimental setup aligns with established DAS benchmark splits \citep{das2010farfar} for compatibility with existing evaluation protocols \citep{joshi2025grnade}. We implement strict length consistency filtering to remove preference pairs where sequence lengths differ or do not match graph node counts, ensuring training stability and preventing spurious length-based correlations. Following preprocessing, our dataset comprises 20,811 training pairs, 505 validation pairs, and 429 test pairs, drawn from 4,025 training instances, 100 validation instances, and 98 test instances.  This testing dataset split has overlap with the training dataset of RiboDiffusion \citep{huang2024ribodiffusion} and RDesign \citep{tanrdesign}, which leads to the abnormally high sequence recovery (100\%) for some structures like 7D7W.

\paragraph{Evaluation Protocol and Metrics}
We evaluate all approaches, generating 8 sequence candidates under temperature $0.1$ per backbone for statistical reliability. We directly apply the SSTT evaluation pipeline (Section\ref{subsec:sstt}) with EternaFold \citep{waymentsteele2022eternafold} for secondary structure predictions, and RhoFold+ \citep{shen2024accurate} for tertiary structure prediction. Furthermore, we report thresholded success rates, including percentages achieving RMSD $\leq$ 15.0\AA, pLDDT $\geq$0.50, MFE $\leq$ -8.0 $kcal/mol$, and TM $\geq$ 0.05. 

We evaluated pass@k performance \citep{lyu2024passk,wu2025invisible} by sampling 64 sequences per RNA structure (folded by RhoFold+ ~\citep{shen2024accurate}) and measuring success rates across different k values (1, 2, 4, 8, 16, 32, 64). For a given structure and k value, pass@k is defined as:
\begin{equation}
\text{pass@k} \;=\; 1-(1-p)^k .
\end{equation}
where p denotes the indicator function, returning 1 when the case pass the criteria, and returning 0 otherwise.

\paragraph{Baseline Approaches}
We compare RiboPO against recent deep learning based RNA inverse folding models spanning two main paradigms. 
\textbf{Autoregressive models} include gRNAde \citep{joshi2025grnade}, RDesign \citep{tanrdesign}, R3Design \citep{tan2025r3design}, RiFold \citep{liu2025sentences}, and RhoDesign \citep{wong2024deep}.
\textbf{Diffusion-based models} include RiboDiffusion \citep{huang2024ribodiffusion}, RhoDesign \citep{wong2024deep} and \textbf{RIdiffusion} \citep{Hou2025ridiffusion}.

\subsection{SSTT Benchmark Analysis}
\label{subsec:benchmark_analysis}

\textbf{Baseline overview.}
We evaluate all four SSTT dimensions comprehensively (Table~\ref{tab:main_results_full}). Among baselines, \textbf{gRNAde} delivers the strongest sequence- and secondary-structure metrics, reflecting its GNN–GVP backbone conditioning. \textbf{Single-round RiboPO improves secondary and thermodynamic properties.}
Our first-round DPO (\textbf{RiboPO Round 1}) shows that even a single training round markedly enhances secondary-structure and thermodynamic properties over the gRNAde base model. Despite a minor decline in recovery (0.5288 $\rightarrow$ 0.504), RiboPO increases diversity (0.83 $\rightarrow$ \textbf{0.90}) and improves scMCC (0.60 $\rightarrow$ 0.71) and MFE ($-32.83$ $\rightarrow$ \textbf{$-35.83$}). This indicates that policy optimization effectively trades off marginal sequence similarity for substantially better foldability and energetic favorability. This represents early-stage exploration of a broader biophysically plausible sequence with improved stability. \textbf{Progressive refinement through multiple rounds.}
As RiboPO proceeds to multiple rounds, \textbf{thermostability gains become striking}. By Round 2, MFE further improves to \textbf{$-36.86$ kcal/mol}, the best of all models, while scMCC remains high (0.70) and tertiary metrics (RMSD and \%RMSD$\leq$8Å) also improve. These gains demonstrate that decreasing $\gamma_r$ across rounds—our “coarse-to-fine” margin schedule—admits increasingly subtle yet energetically favorable sequences, allowing the model to internalize thermodynamic regularities and better capture RNA structure–stability trade-offs. \textbf{Balanced sequence and structural fidelity.}
Although RiboPO prioritizes stability and secondary structure, its tertiary metrics remain competitive. Across all rounds, pLDDT holds steady at $\sim$0.65, RMSD stays near 10.4Å, and TM-score remains comparable to or better than several baselines. This suggests that thermodynamic optimization does not compromise 3D fidelity; instead, multi-round training preserves global structure while refining local and energetic features. In RNA terms, the model learns not only to produce sequences that fold correctly but also to maintain a realistic conformational ensemble.

\begin{table*}[ht!]
\centering
\caption{Main benchmark results on the DAS test set. RiboPO significantly outperforms baselines on thermostability and secondary structure metrics (T and S), while remaining competitive on sequence and tertiary structure metrics (S and T). Best results are in \textbf{bold}.}
\label{tab:main_results_full}
\resizebox{\textwidth}{!}{%
\begin{tabular}{lccccccccc}
\toprule
\multirow{2}{*}{\textbf{Model}} 
& \multicolumn{2}{c}{\cellcolor{blue!10}\textbf{S (Sequence)}} 
& \multicolumn{1}{c}{\cellcolor{blue!10}\textbf{S (2D Struct.)}} 
& \multicolumn{5}{c}{\cellcolor{blue!10}\textbf{T (3D Struct.)}} 
& \multicolumn{1}{c}{\cellcolor{blue!10}\textbf{T (Thermo.)}} \\
\cmidrule(lr){2-3} \cmidrule(lr){4-4} \cmidrule(lr){5-9} \cmidrule(lr){10-10}
& \makecell{Rec.\\($-$)} & \makecell{Div.\\($\uparrow$)} & \makecell{scMCC\\($\uparrow$)} & \makecell{pLDDT\\($\uparrow$)} & \makecell{\%pLDDT\\$\geq$0.70} & \makecell{RMSD\\($\downarrow$)} & \makecell{\%RMSD\\$\leq$8\AA} & \makecell{TM-sc.\\($\uparrow$)} & \makecell{MFE\\($\downarrow$)} \\
\midrule
RiFold & 0.416 & 0.89 & 0.28 & 0.50 & 0.012 & 17.06 & 0.20 & 0.12 & -11.34 \\
RDesign & 0.415 & 0.84 & 0.20 & 0.45 & 0.22 & 16.81 & 0.12 & 0.12 & -10.59 \\
RIDiffusion & \textbf{0.533} & 0.85 & 0.59 & 0.60 & 0.28 & 10.66 & 0.46 & 0.25 & -28.81 \\
gRNAde (Base) & 0.529 & 0.83 & 0.60 & 0.62 & 0.38 & 11.45 & 0.42 & 0.29 & -32.83 \\
\midrule
RiboPO (Round 1) & 0.504 & \textbf{0.90} & 0.71 & \textbf{0.66} & \textbf{0.45} & 10.40 & 0.45 & 0.30 & -35.83 \\
RiboPO (Round 2) & 0.502 & 0.88 & 0.70 & 0.65 & 0.45 & \textbf{10.23} & \textbf{0.51} & \textbf{0.31} &  \textbf{-36.86} \\
RiboPO (Round 4) &  0.506 &  0.89 &  \textbf{0.72} & 0.65 & 0.42 & 10.42 & 0.44 & 0.29 &  -35.22 \\   
\bottomrule
\end{tabular}%
}
\raggedright

\end{table*}

\subsection{Advanced Analysis of Reinforcement Learning Effects}

\begin{figure*}[t!]
\centering

\begin{subfigure}[b]{0.48\textwidth}
\centering
\includegraphics[width=\textwidth]{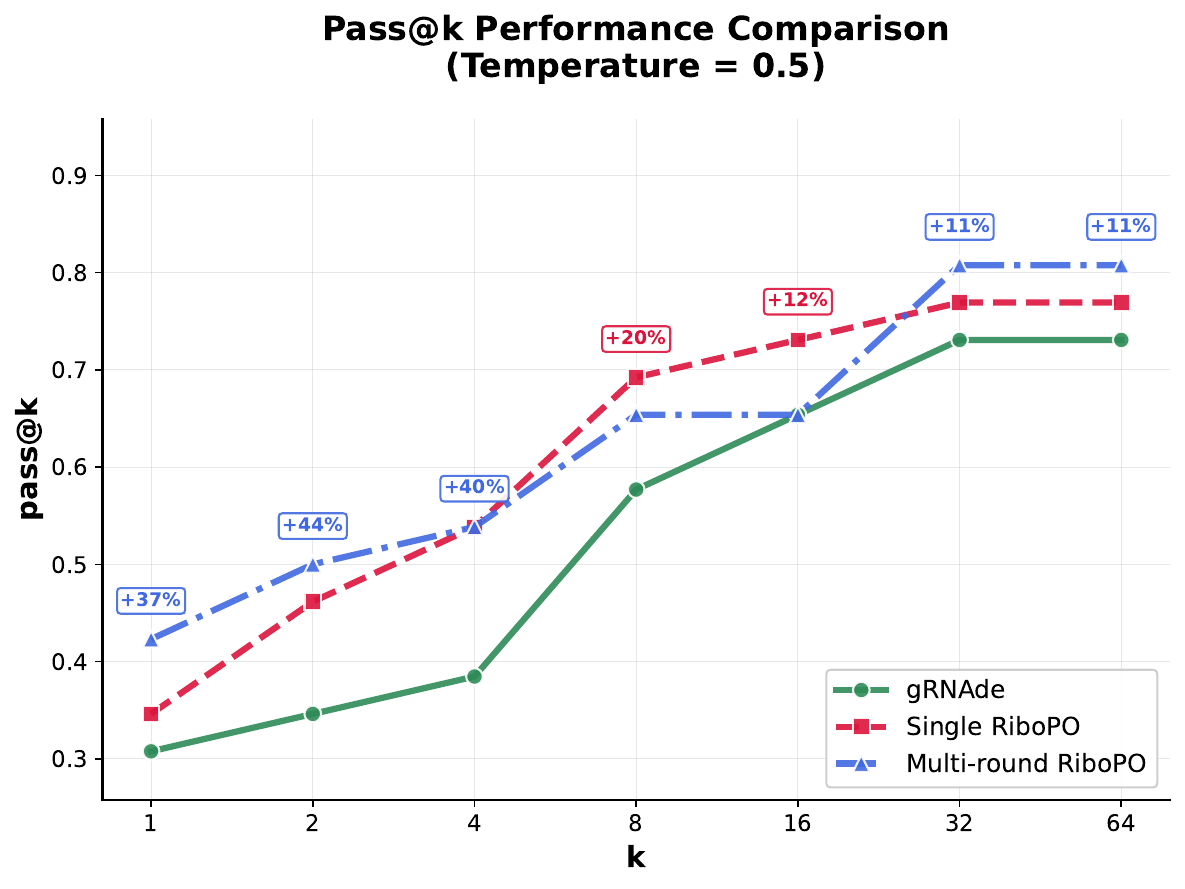}
\caption{Pass@k Sampling Efficiency}
\label{fig:passk_curve}
\end{subfigure}
\hspace{0.02cm}
\begin{subfigure}[b]{0.48\textwidth}
\centering
\includegraphics[width=\textwidth]{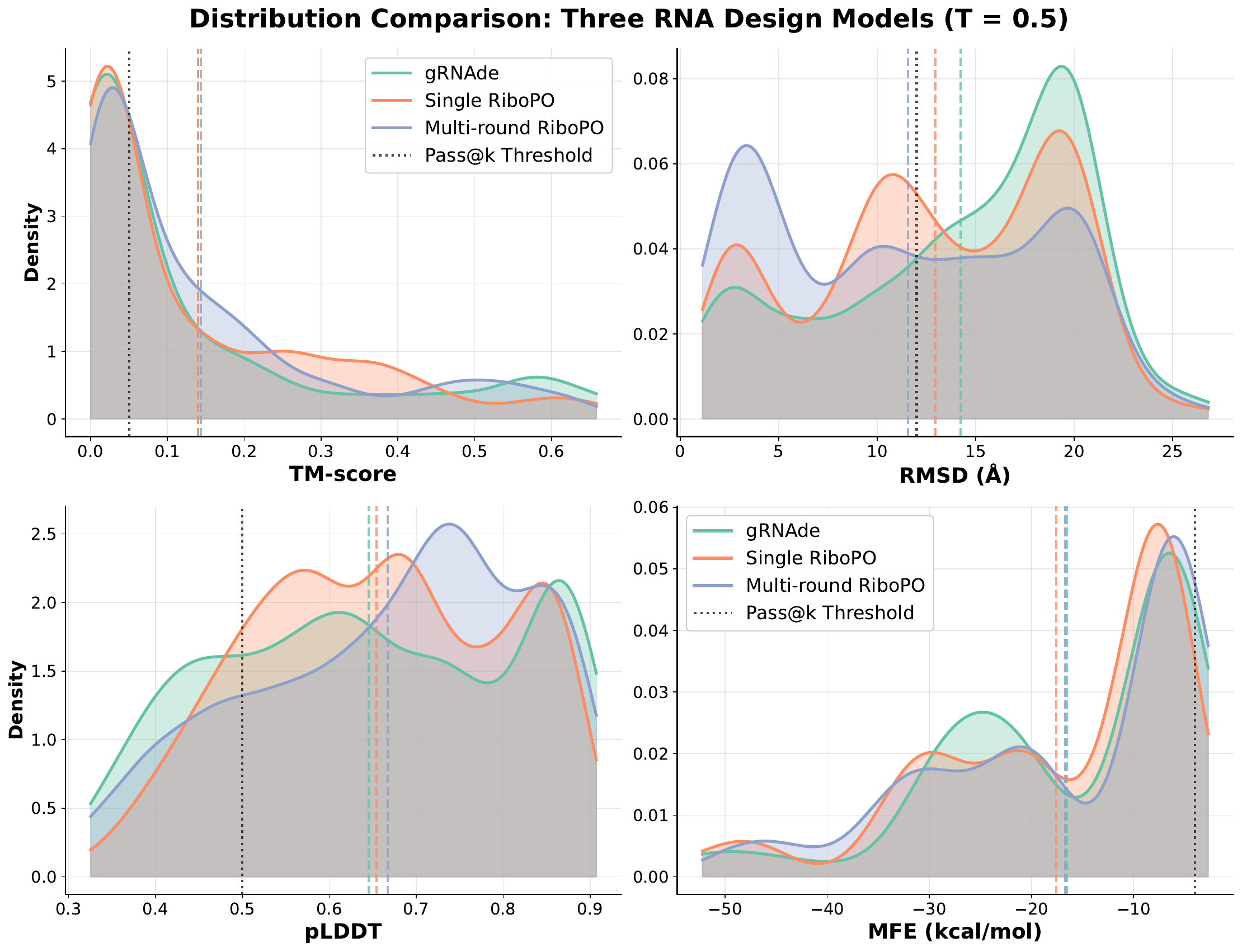}
\caption{Metric Distribution Shifts}
\label{fig:dist_shift}
\end{subfigure}

\vspace{-6pt}
\caption{
\textbf{RiboPO improves both sampling efficiency and the quality distribution of generated sequences.}
\textbf{(a)} Pass@k performance shows RiboPO consistently dominates gRNAde across all sample sizes $k$, with the largest gains at the small-$k$ screening regime crucial for practical applications.
\textbf{(b)} Kernel density estimates for key metrics on 14 held-out RNA structures. Multi-round RiboPO shifts the distributions for RMSD and MFE towards more favorable values (indicated by vertical lines) compared to the gRNAde baseline, demonstrating improved designability and thermostability.
}
\label{fig:combined_analysis}
\vspace{-6pt}
\end{figure*}

\paragraph{RiboPO improves hit rate and sampling efficiency.}
We assess sampling efficiency using pass@k analysis on the identified 14 important RNA structures by \citep{das2010farfar} (Table~\ref{tab:pass_at_k}). For small $k$ values (pass@1–4), RiboPO achieves 37\%-44\% relative increase in success rate. Even at large $k$ (pass@32–64) Figure~\ref{fig:passk_curve}, RiboPO sustains 11\% higher success rates. This demonstrates that DPO improves the “hit rate” of high-quality sequences per sampling budget, which is critical when each candidate must be computationally folded. From an RL perspective, this reflects curriculum-based refinement that concentrates probability mass in high-reward regions; from an RNA perspective, it means fewer trials are needed to find sequences satisfying structural and energetic constraints, thereby reducing in silico or wet-lab burden.


\paragraph{Multi-round DPO drives measurable distribution shifts.}
Figure~\ref{fig:dist_shift} compares gRNAde and RiboPO after multi-round training on TM-score, RMSD, pLDDT, and MFE. We observe that RMSD and MFE distributions move toward higher-quality regions: the incidence of poor outliers (RMSD $>15$\AA, MFE $>-10$ kcal/mol) decreases while the mass of stable, low-energy designs increases (mean MFE $-16.5\rightarrow-18.9$ kcal/mol). pLDDT, already relatively high for the base model (mean $\sim 0.62$), becomes more stable with enrichment in the 0.65–0.8 range and suppression of low-confidence cases. TM-score exhibits a “center-convergence” pattern: scores at the low end are lifted, yet samples concentrate near the middle (0.15–0.40), consistent with indirect optimization via RMSD and pLDDT rather than TM-score itself. From the RL viewpoint, these shifts confirm that large early margins ($\gamma_r$ high) promote coarse exploration and smaller later margins refine toward explicitly rewarded regions (RMSD, MFE). From the RNA viewpoint, the policy first secures gross foldability and energetic feasibility, then favors finer conformational improvements; \textbf{this two-phase process mirrors natural RNA evolution}.

\begin{table}[!t]
\centering
\caption{Pass@k results for generating a sequence with RMSD $< 12$ \AA, pLDDT $> 0.50$, TM-score $> 0.05$, and MFE $< -4.0$ kcal/mol (T = 0.5). RiboPO demonstrates superior sampling efficiency with Multi-round RiboPO achieving the best performance.}
\label{tab:pass_at_k}
\resizebox{\columnwidth}{!}{%
\begin{tabular}{lccccccc}
\toprule
\textbf{Model} & \textbf{pass@1} & \textbf{pass@2} & \textbf{pass@4} & \textbf{pass@8} & \textbf{pass@16} & \textbf{pass@32} & \textbf{pass@64} \\
\midrule
gRNAde & 0.308 & 0.346 & 0.385 & 0.577 & 0.654 & 0.731 & 0.731 \\
RiboPO (Round 1) & 0.346 & 0.462 & 0.538 & \textbf{0.692} & \textbf{0.731} & 0.769 & 0.769 \\
RiboPO (Round 4) & \textbf{0.423} & \textbf{0.500} & \textbf{0.538} & 0.654 & 0.654 & \textbf{0.808} & \textbf{0.808} \\
\bottomrule
\end{tabular}}
\end{table}

\subsection{Ablation Studies}
\label{subsec:ablation_studies}

\paragraph{Ablation on loss components and hyperparameters.}
To test whether our loss formulation truly supports multi-objective optimization, we ablate both loss components and loss weights (Table~\ref{tab:ablation_loss}). Removing the DPO term ($\beta{=}0$) causes broad degradation in structural metrics (pLDDT $0.62$, RMSD $11.50$), confirming that pairwise preference learning is the main driver of structural improvements. Ablating the SFT term ($\lambda_{\text{SFT}}{=}0$) increases diversity ($0.93$) and makes MFE more negative ($-42.60$~kcal/mol), but markedly worsens tertiary quality (TM-score $0.19$, RMSD $13.55$). This indicates that SFT functions as a distributional anchor that prevents over-exploitation of low-energy but geometrically unrealistic regions; \textbf{removing SFT trades stability for 3D realism}. 

Loss-weight ablation further shows a clear trade-off governed by $\beta$. According to the DPO objective, a smaller $\beta$ emphasizes deviation from the reference policy $\pi_{ref}$, thereby pushing the model more aggressively toward aligning with the chosen responses. Consequently, the setting $\beta=0.12$ achieves highly optimized performance across most evaluation metrics. By contrast, $\beta=0.5$ yields slightly better performance than $\beta=1.0$, reflecting a more balanced optimization between fidelity to $\pi_{ref}$ and adaptation to preferences. In the case of MFE, the performance exhibits an inverse relationship with $\beta$: larger values consistently result in superior outcomes. We attribute this to the construction of the pairwise preference data, which introduces no explicit margin for MFE, thereby weakening the available learning signal. In contrast, pLDDT and RMSD provide stronger and more discriminative gradients that dominate optimization when $\beta$ is small, leading to poorer MFE. At larger $\beta$ values, the optimization is more conservative, preserving information from $\pi_{ref}$ and thus mitigating the imbalance among different supervision signals. This explains why MFE benefits disproportionately from larger $\beta$, even though other structural metrics may plateau or decline. These observations highlight that the distribution of the preference dataset plays a pivotal role in shaping DPO performance, which is consistent with recent findings in \cite{Pan2025WhatMI}.


\begin{table}[!t]
\centering
\caption{Ablation study on loss components for single round training}
\label{tab:ablation_loss}
\resizebox{\columnwidth}{!}{%
\begin{tabular}{lccccccccc}
\toprule
\multirow{2}{*}{\textbf{Configuration}} 
& \multicolumn{2}{c}{\cellcolor{blue!10}\textbf{S (Sequence)}} 
& \multicolumn{1}{c}{\cellcolor{blue!10}\textbf{S (2D Struct.)}} 
& \multicolumn{5}{c}{\cellcolor{blue!10}\textbf{T (3D Struct.)}} 
& \multicolumn{1}{c}{\cellcolor{blue!10}\textbf{T (Thermo.)}} \\
\cmidrule(lr){2-3} \cmidrule(lr){4-4} \cmidrule(lr){5-9} \cmidrule(lr){10-10}
& \makecell{Rec.\\($-$)} & \makecell{Div.\\($\uparrow$)} & \makecell{scMCC\\($\uparrow$)} & \makecell{pLDDT\\($\uparrow$)} & \makecell{\%pLDDT\\$\geq$0.70} & \makecell{RMSD\\($\downarrow$)} & \makecell{\%RMSD\\$\leq 8\,\text{\AA}$} & \makecell{TM-sc.\\($\uparrow$)} & \makecell{MFE\\($\downarrow$)} \\
\midrule
w/o $\mathcal{L}_{\mathrm{DPO}}$ ($\beta=0$) & 0.49 & 0.85 & 0.61 & 0.62 & 0.36 & 11.50 & 0.36 & 0.27 & -33.68 \\
w/o $\mathcal{L}_{\mathrm{SFT}}$ ($\lambda_{SFT}=0$) & 0.44 & 0.93 & 0.64 & 0.62 & 0.42 & 13.55 & 0.39 & 0.19 & \textbf{-42.60} \\
\midrule
RiboPO ($\beta=0.5$) & 0.47 & 0.88 & 0.57 & 0.62 & 0.40 & 11.96 & 0.45 & 0.25 & -41.05 \\
RiboPO ($\beta=1.0$) & 0.49 & 0.85 & 0.59 & 0.60 & 0.35 & 13.90 & 0.26 & 0.20 & -41.86 \\
\midrule
RiboPO ($\beta=0.12$) & \textbf{0.50} & \textbf{0.90} & \textbf{0.71} & \textbf{0.66} & \textbf{0.45} & \textbf{10.40} & \textbf{0.45} & \textbf{0.30} & -35.83 \\
\bottomrule
\end{tabular}%
}
\end{table}

\paragraph{Ablation on offline RL algorithms.}
We compare SimPO and DPO under a single round (Table~\ref{tab:ablation_algo}). DPO consistently outperforms SimPO on structural fidelity (scMCC $0.71$ vs.\ $0.66$, RMSD $10.40$ vs.\ $11.36$, TM-score $0.30$ vs.\ $0.27$) and on robustness thresholds (\%RMSD$\leq$8\AA: $0.45$ vs.\ $0.40$; \%pLDDT$\geq 0.70$: $0.45$ vs.\ $0.44$), while maintaining higher diversity ($0.90$ vs.\ $0.87$). SimPO shows slightly higher recovery ($0.51$ vs.\ $0.50$) and more negative MFE ($-36.28$ vs.\ $-35.83$). These results suggest that DPO’s likelihood-ratio objective with an explicit reference anchor is better at preserving global 3D realism while still exploring diverse sequences; \textbf{SimPO favors stability and exact-token matching slightly more, but at the expense of tertiary structure}.

\begin{table}[ht!]
\centering
\caption{Comparison of different RL algorithms (DPO and SimPO) with only a single round. DPO demonstrates superior performance across key metrics.}
\label{tab:ablation_algo}
\resizebox{\columnwidth}{!}{%
\begin{tabular}{lccccccccc}
\toprule
\multirow{2}{*}{\textbf{Algorithm}} 
& \multicolumn{2}{c}{\cellcolor{blue!10}\textbf{S (Sequence)}} 
& \multicolumn{1}{c}{\cellcolor{blue!10}\textbf{S (2D Struct.)}} 
& \multicolumn{5}{c}{\cellcolor{blue!10}\textbf{T (3D Struct.)}} 
& \multicolumn{1}{c}{\cellcolor{blue!10}\textbf{T (Thermo.)}} \\
\cmidrule(lr){2-3} \cmidrule(lr){4-4} \cmidrule(lr){5-9} \cmidrule(lr){10-10}
& \makecell{Rec.\\($-$)} & \makecell{Div.\\($\uparrow$)} & \makecell{scMCC\\($\uparrow$)} & \makecell{pLDDT\\($\uparrow$)} & \makecell{\%pLDDT\\$\geq$0.70} & \makecell{RMSD\\($\downarrow$)} & \makecell{\%RMSD\\$\leq$8\AA} & \makecell{TM-sc.\\($\uparrow$)} & \makecell{MFE\\($\downarrow$)} \\
\midrule
SimPO & \textbf{0.51} & 0.87 & 0.66 & 0.65 & 0.44 & 11.36 & 0.40 & 0.27 & \textbf{-36.28} \\
DPO & 0.50 & \textbf{0.90} & \textbf{0.71} & \textbf{0.66} & \textbf{0.45} & \textbf{10.40} & \textbf{0.45} & \textbf{0.30} & -35.83 \\
\bottomrule
\end{tabular}%
}
\end{table}

\paragraph{Analysis of multi-round training strategy.}
Table~\ref{tab:ablation_multiround} disentangles three design choices: curriculum on the preference margin (CL), reference-model updates across rounds, and algorithm family. Multi-round DPO without curriculum already improves over the base, but exhibits moderate RMSD ($13.39$) and limited high-confidence structures (\%pLDDT$\geq 0.70{=}0.39$). Introducing a curriculum that narrows the margin across rounds (\emph{Multi-Round CL DPO}) yields the strongest overall structural performance (scMCC $0.72$, pLDDT $0.65$, RMSD $10.42$, \%RMSD$\leq 8$\AA\ $0.44$, TM-score $0.29$) with good diversity ($0.89$). This aligns with our design: early coarse margins de-noise preferences; later fine margins admit subtle yet consistent improvements, steering the policy into near-native basins. 

Updating the reference model each round dramatically pushes MFE downward ($-48.07$~kcal/mol) but harms structure (RMSD $14.30$, TM-score $0.16$). The moving reference weakens the trust-region effect and effectively “ratchets” the policy toward low-energy pockets even when geometry degrades, indicating reward hacking on stability proxies. A curriculum with SimPO across rounds also achieves very low MFE yet lags on structure, reinforcing that the objective form and anchoring are critical. \textbf{The best balance arises from \emph{fixed-reference} multi-round DPO with a margin curriculum}, which preserves 3D fidelity while improving stability.

\begin{table}[ht!]
\centering
\caption{Ablation study on multi-round training components. Updating the reference model and using a curriculum for the preference margin are both beneficial. MR: Multi-Round; CL: Curriculum Learning, with narrowing preference margin; Take the checkpoint evaluation results at the end of final iteration.}
\label{tab:ablation_multiround}
\resizebox{\columnwidth}{!}{%
\begin{tabular}{lccccccccc}
\toprule
\multirow{2}{*}{\textbf{Strategy}} 
& \multicolumn{2}{c}{\cellcolor{blue!10}\textbf{S (Sequence)}} 
& \multicolumn{1}{c}{\cellcolor{blue!10}\textbf{S (2D Struct.)}} 
& \multicolumn{5}{c}{\cellcolor{blue!10}\textbf{T (3D Struct.)}} 
& \multicolumn{1}{c}{\cellcolor{blue!10}\textbf{T (Thermo.)}} \\
\cmidrule(lr){2-3} \cmidrule(lr){4-4} \cmidrule(lr){5-9} \cmidrule(lr){10-10}
& \makecell{Rec.\\($-$)} & \makecell{Div.\\($\uparrow$)} & \makecell{scMCC\\($\uparrow$)} & \makecell{pLDDT\\($\uparrow$)} & \makecell{\%pLDDT\\$\geq$0.70} & \makecell{RMSD\\($\downarrow$)} & \makecell{\%RMSD\\$\leq$8\AA} & \makecell{TM-sc.\\($\uparrow$)} & \makecell{MFE\\($\downarrow$)} \\
\midrule
Multi-Round DPO, w/o CL & 0.49 & \textbf{0.91} & 0.69 & 0.64 &  0.39 & 13.39  & 0.39 & 0.24 & -35.10 \\
Reference Model Update & 0.47 & 0.85 & 0.63 & 0.60 & 0.35 & 14.30 & 0.17 & 0.16 & -48.07 \\
Multi-Round CL SimPO & 0.46 & 0.87 & 0.61 & 0.58 & 0.25 & 13.14 & 0.32 & 0.21 & \textbf{-48.22} \\
Multi-Round CL DPO & \textbf{0.51} & 0.89 & \textbf{0.72} & \textbf{0.65} & \textbf{0.42} & \textbf{10.42} & \textbf{0.44} & \textbf{0.29} & -35.22 \\
\bottomrule
\end{tabular}%
}
\end{table}



\section{Conclusion}
We introduced \textbf{RiboPO}, a backbone-conditioned preference optimization framework that treats RNA inverse folding as a multi-objective problem over structural fidelity and thermodynamic stability. Preference pairs are built by a feasibility gate (pLDDT/RMSD/MFE) and a round-wise margin, and optimized via DPO with a fixed reference plus an SFT anchor. On SSTT, RiboPO improves thermostability and secondary-structure self-consistency while remaining competitive on tertiary metrics, and raises pass@k hit rates—evidence of more efficient sampling. Ablations show that DPO drives structural gains, SFT controls distributional drift and preserves geometry, and the margin curriculum enables coarse-to-fine refinement.

Our analysis also clarifies limits of offline preference alignment: energy-leaning signals shift the policy toward low-MFE basins (explaining recovery drops), and 2D thermodynamic proxies do not fully align with global 3D similarity under current RNA predictors. Moving forward, we will integrate 3D-aware and verifiable rewards in a semi-online, multi-round scheme and calibrate 2D/3D objectives under a Pareto-aware formulation. \textbf{Overall, preference-based RL reliably steers sequences toward biophysically plausible, designable RNAs while preserving 3D fidelity.}

\bibliography{iclr2026_conference}
\bibliographystyle{iclr2026_conference}

\newpage
\appendix
\section{Appendix}

\subsection{Implementation details.}
\paragraph{Training details.} Our training protocol employs cosine learning rate decay with 1,000-step warmup, gradient accumulation across 4 steps, batch size of 32 per preference pair, and precision bf16 training. Training proceeds for minimum 5 epochs, validation every 50 steps, and model checkpoints saved at the same interval.

\section{SSTT: Comprehensive Fine-Grained RNA Design}
\label{subsec:sstt_detail}

\vspace{0.5ex}
\noindent\textbf{Notation.} For a candidate sequence $y=(y_1,\dots,y_L)$ and the native sequence $y^{\mathrm{nat}}$, let $S_0$ denote the target secondary structure, $U_0$ the set of unpaired positions in $S_0$, and $P_0$ the set of paired positions with partner map $j^*(i)$. Partition-function outputs are base-pair probabilities $p_{ij}$ and unpaired probabilities $q_i$. Predicted 3D coordinates are $\mathbf{X}(y)$; native/target coordinates are $\mathbf{X}^{\mathrm{ref}}$.

\subsection*{Sequence axis (S)}
\begin{itemize}[leftmargin=1.2em,itemsep=2pt,topsep=2pt]
\item \textbf{Recovery} \citep{dauparas2022proteinmpnn}: Recovery measures the average percentage of native nucleotides correctly recovered in the sampled sequences. It is computed as

$$\mathrm{Rec}(y) \;=\; \frac{1}{L} \sum_{i=1}^{L} \mathbb{1}\{y_i = y_i^{\mathrm{nat}}\}.$$

\item \textbf{Perplexity} (↓, lower-is-better) \citep{bengioaneural}: Perplexity measures how well a model predicts the sequences. Given a model (reference or policy) with token probabilities $p_\theta$, perplexity is defined as the average exponential of the negative log-likelihood of the sampled sequences.

$$\mathrm{PPL}(y) \;=\; \exp\!\Big(-\frac{1}{L} \sum_{i=1}^{L} \log p_\theta(y_i \mid y_{<i}, \mathcal{G})\Big).$$

\item \textbf{Diversity (3-mer)} (↑): Given $n$ candidates for the backbone $\mathcal{G}$, let $v_s \in \mathbb{R}^{64}$ be the normalized 3-mer frequency of candidate $s$. Define

$$\mathrm{Div}_{3\mathrm{mer}}(\mathcal{G}) \;=\; 1 \;-\; \frac{2}{n(n-1)} \sum_{1 \le s < t \le n} \rho\!\big(v_s, v_t\big),$$

where $\rho(\cdot,\cdot)$ is the Pearson correlation. It's the quantification of sequence variety based on substring frequency distributions \citep{shannon1948mathematical, Bokulich2024diversitykmer}.
\end{itemize}

\subsection*{Secondary axis (S) — forward-folding self-consistency}
\begin{itemize}[leftmargin=1.2em,itemsep=2pt,topsep=2pt]
\item \textbf{2D (EternaFold) scMCC (Matthews Correlation Coefficient) \citep{matthews1975comparison, mccmoreaccurate}} (\(\uparrow\)): This metric measures the ability of designs to recover base pairing patterns. We use EternaFold \citep{waymentsteele2022eternafold} to perform forward folding on the sampled sequences and then calculate the Matthews correlation coefficient (MCC) \citep{joshi2025grnade} by comparing the predicted base-pair adjacency matrix with the ground truth, using the formula:
\[
\mathrm{scMCC}_{\mathrm{2D}}=\frac{\mathrm{TP}\cdot \mathrm{TN}-\mathrm{FP}\cdot \mathrm{FN}}{\sqrt{(\mathrm{TP}+\mathrm{FP})(\mathrm{TP}+\mathrm{FN})(\mathrm{TN}+\mathrm{FP})(\mathrm{TN}+\mathrm{FN})}},
\]
with the optional F1 score.
\end{itemize}

\subsection*{Thermostability axis (T) — ensemble specificity}
\begin{itemize}[leftmargin=1.2em,itemsep=2pt,topsep=2pt]
\item \textbf{MFE} (kcal/mol, \(\downarrow\)): The minimum free energy (MFE) of the most stable predicted secondary structure. \emph{Note:} MFE is a single-structure quantity and does not capture ensemble specificity; we therefore also report Ensemble defect per nucleotide (ED/nt), Target probability \(P(S_0)\), and Shannon entropy.
\item \textbf{Ensemble defect (ED) \citep{zadeh2011nucleic} and ED/nt} (\(\downarrow\)): The ensemble defect (ED) measures the discrepancy between the predicted base-pair probabilities and the true structural states, including both unpaired and paired positions. The formula is defined as:
\[
\mathrm{ED}(y;S_0)=\sum_{i\in U_0}(1-q_i)+\sum_{i\in P_0}\big(1-p_{i\,j^*(i)}\big),\qquad \mathrm{ED/nt}=\mathrm{ED}/L.
\]

where $U_0$ represents the set of unpaired positions, $P_0$ is the set of paired positions, $q_i$ is the unpaired probability, and $p_{i\,j^*(i)}$ is the base-pairing probability of a paired position.
\item \textbf{Target probability} (\(\uparrow\)):
\[
P(S_0)=\frac{e^{-\beta \Delta G(S_0)}}{Z},\qquad Z=\sum_{S} e^{-\beta \Delta G(S)},\quad \beta=1/(RT).
\]
\end{itemize}

At a given temperature, an RNA molecule exists not as a single static structure but as a thermodynamic ensemble of all possible conformations, each populated according to its Boltzmann weight $e^{-E/k_B T}$ \citep{McCaskill1990TheEquilibrium}. The partition function \citep{mathews2004usingpartition}, denoted by $Z$, is a fundamental quantity in statistical mechanics that encapsulates the properties of this entire ensemble. It is defined as the sum of the Boltzmann weights over all possible secondary structures $S$.




\subsection*{Tertiary axis (T)}
\begin{itemize}[leftmargin=1.2em,itemsep=2pt,topsep=2pt]
\item \textbf{TM-score} \citep{zhang2004tmscore} (↑): The TM-score evaluates the structural alignment quality between the predicted and reference RNA structures after optimal superposition. It is calculated based on the distances $d_i$ between aligned residues and a length-dependent scale $d_0(L)$. The formula for TM-score is given by:

$$
\mathrm{TM}(y) \;=\; \frac{1}{L_{\mathrm{ref}}} \sum_{i=1}^{L_{\mathrm{ali}}} \frac{1}{1 + \left(\frac{d_i}{d_0(L_{\mathrm{ref}})}\right)^2},
\quad
d_0(L) \;=\; 1.24\sqrt[3]{L-15}-1.8.
$$

We report TM-score, GDT (↑), RMSD (↓), and thresholded success slices: \%\,$\mathrm{TM}\ge 0.45$, \%\,$\mathrm{GDT}\ge 0.50$, \%\,$\mathrm{RMSD}\le 8$~\AA, following the previous research \citep{joshi2025grnade}. US-align \citep{zhang2022usalign} is deployed for the calculation of the TM-score.
\item \textbf{pLDDT} (↑): The mean predicted LDDT confidence \citep{jumper2021alphafold2} provides an estimate of the accuracy of local structural predictions. We also report \%\,$\mathrm{pLDDT}\ge 0.70$ \citep{Williams2025plddtcategorizing}.
\item \textbf{lDDT} (↑) and \textbf{lDDT success rate} (↑): The local distance difference test (lDDT) is a metric that evaluates the accuracy of predicted distances between atoms, considering their local structural context. The neighbor set $N_i$ is defined in the \emph{reference} by a cutoff (e.g., 15~\AA) and tolerances $\Delta \in \{0.5,1,2,4\}$~\AA. The lDDT is calculated as follows:

$$\mathrm{lDDT}(y) \;=\; \frac{1}{L} \sum_{i=1}^{L} \frac{1}{|N_i|} \sum_{j \in N_i} \frac{1}{4}\sum_{\Delta} \mathbb{1}\!\left( \left| d_{ij}(y)-d_{ij}^{\mathrm{ref}}\right| < \Delta \right).$$

\item \textbf{INF (ALL/WC/NWC/STACK) \citep{parisien2009newmetricsinf}} (↑): Interaction-network fidelity (INF) is used to assess the accuracy of predicted RNA interactions by comparing them with the reference interaction sets. It is computed as MCC (or F1) between the predicted and reference interaction sets. Let $(\mathrm{TP},\mathrm{FP},\mathrm{FN},\mathrm{TN})$ be counts over the \emph{union} of interactions under consideration (e.g., ALL = WC$\cup$NWC$\cup$STACK). Then

$$
\mathrm{INF}_{\bullet}^{\mathrm{MCC}} \;=\;
\frac{\mathrm{TP}\cdot \mathrm{TN} - \mathrm{FP}\cdot \mathrm{FN}}{\sqrt{(\mathrm{TP}+\mathrm{FP})(\mathrm{TP}+\mathrm{FN})(\mathrm{TN}+\mathrm{FP})(\mathrm{TN}+\mathrm{FN})}}.
$$

Note that $\mathrm{INF}_{\mathrm{ALL}}$ is \emph{not} the sum of per-class scores; it is computed on the pooled confusion matrix.
\item \textbf{Clash score} ($\downarrow$): Clash score measures the number of all-atom steric clashes per 1000 atoms \citep{word1999clashscore, davis2007molprobity}, reported \emph{without} relaxation.
\item \textbf{MCQ torsion metrics} ($\downarrow$/$\uparrow$/$\downarrow$): The MCQ torsion metrics are used to evaluate the accuracy of backbone torsion angles in the predicted structure. let $\theta_i^{\mathrm{ref}}$ and $\theta_i$ be corresponding backbone torsions. Define circular errors $\delta_i = \mathrm{angdiff}(\theta_i,\theta_i^{\mathrm{ref}}) \in (-\pi,\pi]$. Then

$$
\mathrm{MCQ}_{\mathrm{abs}} \;=\; \frac{180}{\pi} \cdot \frac{1}{K} \sum_{i=1}^{K} |\delta_i|,\quad
R \;=\; \Big\| \frac{1}{K}\sum_{i=1}^{K} e^{\mathrm{i}\delta_i} \Big\|,\quad
\sigma_{\circ} \;=\; \frac{180}{\pi}\sqrt{-2\ln R},
$$

where $K$ is the number of defined torsions \citep{ramachandran1963stereochemistry}; we report $\mathrm{MCQ}_{\mathrm{abs}}$ (↓), $R$ (↑), and $\sigma_{\circ}$ (↓).
\end{itemize}

\subsection*{Implementation notes and recommendations}
\begin{itemize}[leftmargin=1.2em,itemsep=2pt,topsep=2pt]
\item \emph{Thermostability reporting.} Because MFE is single-structure and ED/nt is ensemble-specificity, we always report both; we add $P(S_0)$ to capture the sharpness of the ensemble.
\item \emph{INF family.} We use $\mathrm{INF}_{\mathrm{ALL}}$ as the primary contact metric in the main tables and include WC/NWC/STACK breakdowns in the appendix; negative $\mathrm{INF}_{\mathrm{NWC}}$ can occur under MCC (sparse non-canonicals), which is expected and informative.
\item \emph{MCQ.} We report the triplet $(\mathrm{MCQ}_{\mathrm{abs}}, R, \sigma_{\circ})$ to assess backbone realism—orthogonal to TM/RMSD/INF.
\item \emph{Diversity.} The 3-mer correlation-based diversity emphasizes distributional breadth across samples for the same backbone; when comparing models, we match $n$ and temperature $\tau$.
\end{itemize}

\section{Full RiboPO SSTT Results and Baselines (Extended)}
\label{sec:full_results_appendix}

\begin{table}[H]
\centering
\caption{Sequence and secondary-structure metrics (SSTT: S + S).}
\label{tab:seq_secondary}
\begin{tabular}{lccc}
\toprule
Model & Recovery (-) & Diversity (3-mer) (↑) & scMCC (Eterna) (↑) \\
\midrule
RiFold                      & 0.416 & 0.89 & 0.281 \\
RDesign                     & 0.415 & 0.84 & 0.197 \\
R3Design\textsuperscript{*} & 0.539 & 0.41 & 0.663 \\
RiboDiffusion\textsuperscript{*} & 0.775 & 0.22 & 0.858 \\
RhoDesign\textsuperscript{*} & 0.648 & 0.62 & 0.294 \\
RIDiffusion                 & \textbf{0.533} & 0.85 & 0.585 \\
gRNAde (Base)               & 0.529 & 0.82 & 0.606 \\
\midrule
RiboPO Round 1              & 0.504 & \textbf{0.90} & 0.706 \\
RiboPO Round 2              & 0.502 & 0.87 & 0.697 \\
RiboPO Round 4              & 0.506 & 0.89 & \textbf{0.715} \\
\bottomrule
\end{tabular}
\par\small\emph{* marks baselines whose test sets overlap with the DAS split used here; their strong numbers should be interpreted with caution.}
\end{table}

\begin{table}[H]
\centering
\caption{Tertiary alignment and thresholded success slices (global metrics).}
\label{tab:tertiary_global}
\resizebox{\textwidth}{!}{%
\begin{tabular}{lcccccc}
\toprule
Model & TM (↑) & \% TM $\geq$ 0.45 (↑) & GDT (↑) & \% GDT $\geq$ 0.50 (↑) & RMSD (\AA) (↓) & \% RMSD $\leq$ 8\AA\ (↑) \\
\midrule
RiFold                      & 0.122 & 0.024 & 0.121 & 0.012 & 17.06 & 0.205 \\
RDesign                     & 0.120 & 0.012 & 0.124 & 0.024 & 16.81 & 0.217 \\
R3Design\textsuperscript{*} & 0.481 & 0.578 & 0.470 & 0.554 &  6.20 & 0.759 \\
RiboDiffusion\textsuperscript{*} & 0.618 & 0.838 & 0.633 & 0.868 &  3.69 & 0.904 \\
RhoDesign\textsuperscript{*} & 0.264 & 0.250 & 0.272 & 0.263 & 12.17 & 0.395 \\
RIDiffusion                 & 0.250 & 0.168 & 0.272 & 0.157 & 10.66 & 0.457 \\
gRNAde (Base)               & 0.293 & 0.301 & 0.285 & 0.286 & 11.44 & 0.418 \\
\midrule
RiboPO Round 1              & 0.299 & 0.302 & 0.291 & 0.273 & 10.40 & 0.453 \\
RiboPO Round 2              & \textbf{0.309} & \textbf{0.325} & \textbf{0.314} & \textbf{0.301} & \textbf{10.23} & \textbf{0.505} \\
RiboPO Round 4              & 0.289 & 0.287 & 0.280 & 0.239 & 10.42 & 0.443 \\
\bottomrule
\end{tabular}%
}
\par\small\emph{* marks baselines whose test sets overlap with the DAS split used here; their strong numbers should be interpreted with caution.}
\end{table}

\begin{table}[H]
\centering
\caption{Confidence and local-geometry metrics.}
\label{tab:tertiary_local}
\begin{tabular}{lccc}
\toprule
Model & pLDDT (↑) & \% pLDDT $\geq$ 0.70 (↑) & lDDT (↑) \\
\midrule
RiFold                      & 0.502 & 0.012 & 0.149 \\
RDesign                     & 0.453 & 0.024 & 0.147 \\
R3Design\textsuperscript{*} & 0.678 & 0.542 & 0.183 \\
RiboDiffusion\textsuperscript{*} & 0.768 & 0.824 & 0.193 \\
RhoDesign\textsuperscript{*} & 0.559 & 0.289 & 0.150 \\
RIDiffusion                 & 0.609 & 0.282 & \textbf{0.168} \\
gRNAde (Base)               & 0.621 & 0.381 & 0.141 \\
\midrule
RiboPO Round 1              & \textbf{0.655} & \textbf{0.450} & 0.142 \\
RiboPO Round 2              & 0.652 & 0.449 & 0.143 \\
RiboPO Round 4              & 0.646 & 0.424 & 0.141 \\
\bottomrule
\end{tabular}
\par\small\emph{* marks baselines whose test sets overlap with the DAS split used here; their strong numbers should be interpreted with caution.}
\end{table}

\begin{table}[H]
\centering
\caption{Contact fidelity and stereochemical quality.}
\label{tab:contacts_stereo}
\resizebox{\textwidth}{!}{%
\begin{tabular}{lcccccccc}
\toprule
Model & INF (ALL) (↑) & INF (WC) (↑) & INF (non-WC) (↑) & INF (STACK) (↑) & Clash score (no relax) (↓) & MCQ$_{\mathrm{abs}}$ (°) (↓) & MCQ $R$ (↑) & MCQ $\sigma$ (°) (↓) \\
\midrule
RiFold & 0.39 & -0.06 & -0.07 & 0.55 & 678.3 & 40.73 & 0.64 & 54.52 \\
RDesign & 0.38 & -0.08 & -0.06 & 0.54 & 740.6 & 41.11 & 0.63 & 54.80 \\
R3Design\textsuperscript{*} & 0.58 & 0.34 & 0.00 & 0.73 & 569.5 & 29.16 & 0.77 & 40.31 \\
RiboDiffusion\textsuperscript{*} & 0.65 & 0.49 & 0.00 & 0.78 & 504.7 & 30.18 & 0.76 & 42.02 \\
RhoDesign\textsuperscript{*} & 0.49 & 0.13 & 0.01 & 0.65 & 685.0 & 37.61 & 0.67 & 50.80 \\
RIDiffusion & \textbf{0.52} & \textbf{0.29} & \textbf{-0.01} & \textbf{0.66} & 644.0 & 37.58 & 0.67 & 51.22 \\
gRNAde (Base) & 0.48 & 0.10 & -0.07 & 0.63 & 637.5 & \textbf{33.81} & \textbf{0.72} & \textbf{46.23} \\
\midrule
RiboPO Round 1 & 0.50 & 0.15 & -0.04 & 0.65 & \textbf{595.5} & 34.04 & 0.71 & 46.86 \\
RiboPO Round 2 & 0.49 & 0.15 & -0.05 & 0.64 & 617.9 & 33.85 & 0.72 & 46.59 \\
RiboPO Round 4 & 0.49 & 0.13 & -0.06 & 0.64 & 613.6 & 34.37 & 0.71 & 47.23 \\
\bottomrule
\end{tabular}%
}
\par\small\emph{* marks baselines whose test sets overlap with the DAS split used here; their strong numbers should be interpreted with caution.}
\end{table}

\begin{table}[H]
\centering
\caption{Thermodynamics and ensemble-specificity metrics.}
\label{tab:thermo}
\begin{tabular}{lccc}
\toprule
Model & MFE (kcal/mol) ($\downarrow$) & Ensemble defect / nt ($\downarrow$) & $P(S_0)$ ($\uparrow$) \\
\midrule
RiFold & -11.35 & 0.0034 & 0.0127 \\
RDesign & -10.59 & 0.0036 & 0.0153 \\
R3Design\textsuperscript{*} & -24.49 & 0.0023 & 0.0434 \\
RiboDiffusion\textsuperscript{*} & -22.40 & 0.0025 & 0.1484 \\
RhoDesign\textsuperscript{*} & -15.76 & 0.0069 & 0.4908 \\
RIDiffusion & -28.68 & 0.0626 & \textbf{0.0835} \\
gRNAde (Base) & -16.75 & 0.0028 & 0.0013 \\
\midrule
RiboPO Round 1 & -35.83 & \textbf{0.0021} & 0.0077 \\
RiboPO Round 2 & \textbf{-36.86} & 0.0043 & 0.0203 \\
RiboPO Round 4 & -35.22 & 0.0043 & 0.0122 \\
\bottomrule
\end{tabular}
\par\small\emph{* marks baselines whose test sets overlap with the DAS split used here; their strong numbers should be interpreted with caution.}
\end{table}

\end{document}